\documentstyle[12pt]{article}
\newcommand{\Ibb}[1]{ {\rm I\ifmmode\mkern
	    -3.6mu\else\kern -.2em\fi#1}}
\newcommand{\ibb}[1]{\leavevmode\hbox{\kern.3em\vrule
     height 1.2ex depth -.3ex width .2pt\kern-.3em\rm#1}}
\newcommand{\Cx}{{\ibb C}}

\newcommand{\Rl}{{\Ibb R}}
\newcommand{\Nl}{{\Ibb N}}
\renewcommand{\thefootnote}{\fnsymbol {footnote}}

\newcommand{\n}{\noindent}

\newcommand{\tini}{\longrightarrow}
\newcommand{\slim}{\mathop{\mbox{s--lim}}}

\begin{document}

\baselineskip=18pt
\vskip2truecm
\begin{center}
{\large CLASSICAL MARKOVIAN KINETIC EQUATIONS: 
 EXPLICIT FORM AND H-THEOREM}\\

\vskip0.8cm
 Constantinos Tzanakis\\ 
University of Crete, 74100 Rethymnon, Crete, Greece \\
\vskip0.5cm
 Alkis P. Grecos\footnote[2]{Association Euratom-Etat Belge} \\
Service de Physique Statistique et plasmas \\
 Universit\'{e} Libre de Bruxelles, Campus Plaine (CP.231)\\ 
1050 Bruxelles, Belgium\\ 
\vskip0.8cm

{\sc Abstract}\\
\vskip0.4cm
\end{center}  

{\footnotesize

The probabilistic description of finite classical systems often
leads to linear kinetic equations. A set of physically
motivated mathematical requirements is accordingly formulated.
We show that it necessarily implies  that solutions of such a
kinetic equation in the Heisenberg representation,  define
Markov semigroups on the space of observables. Moreover, a
general $H$-theorem for the adjoint of such semigroups is formulated and
proved provided that at least locally, an invariant measure exists. Under a
certain continuity assumption, the Markov  semigroup property
is sufficient for a linear kinetic equation to be a second
order differential equation with nonegative-definite leading
coefficient. Conversely it is shown that such equations define
Markov semigroups satisfying an $H$-theorem, provided  there exists a
nonnegative equilibrium solution for their formal adjoint, vanishing
at infinity.}

\vskip 0.9truecm
\renewcommand{\thefootnote}{\arabic {footnote}}
\setcounter{footnote}{0}

\centerline{\bf 1. INTRODUCTION}
\bigskip

In a probabilistic description of classical systems with a
finite number of degrees of freedom, one often encounters
{\em linear, autonomous,  evolution equations} of the form
$$
{\partial f \over \partial t}=Z_S f. \eqno(1.1)
$$
\n Such equations, which henceforth, and for the sake of
brevity, will be called kinetic equations, are assumed
to determine the time evolution of the statistical state of the
system. In this context, {\em states are taken to be probability
densities on the phase space}, the manifold of the dynamical
variables characterizing the system, while observables $A$ are
simply (sufficiently well-behaved, real)  phase-space
functions. Furthermore, expectation values are obtained via
some  bilinear form~$w$ on the cartesian product of the spaces
of states and  observables
$$
\langle A \rangle_f \equiv w(f,A), \eqno(1.2)
$$
\n provided they remain finite for all observables, 
i.e. $|w|<+\infty$. If the evolution of  
observables is considered, an equation of the form
$$
{\partial A \over \partial t}=Z_O A, \eqno(1.1') 
$$
\n should hold, with $Z_O$ a formal adjoint in the sense
that
$$
w(f,Z_O A)=w(Z_S f,A) \eqno(1.3)
$$
\n Here it is being assumed that $w$ separates the  states and 
observables (equivalently: $w$ is nondegenerate) and 
$$
w(f_t , A) = w(f , A_t)
$$
\n  i.e.  the two pictures are essentially equivalent.
Typical examples are well known kinetic equations, like those of
Kramers, Fokker-Planck or the (linear) Boltzmann and Landau
equations$^{1-3}$. Such equations result by applying some more or
less systematic approximation scheme based on a perturbation analysis
of the exact microscopic dynamics of the system under
consideration$^{3, 4}$,
or by applying stochastic methods$^{2, 5}$, based on assumptions
concerning the behaviour of a large number of microscopic events
characterizing the system.
For classical open systems weakly interacting with large equilibrium
baths, kinetic equations have been derived$^{6, 7}$ using techniques
similar to those developped for quantum systems$^8$. On the other hand,
some `markovianization' procedures imposed on the exact dynamics may
lead to inconsistent equations$^9$. Note that in such derivations,
expansions with respect to some appropriate parameter are used,
implying that $Z_S$ or $Z_O$ are differential operators of some
order, depending on the approximation.  

However in the physics litterature, not sufficient
attention has been given  to a general
 presentation of the structure and properties of
such {\em classical} (``Markovian") kinetic equations, {\em independently
of the method used to derive them}$^{10}$. In our opinion this seems
necessary at least for two reasons :

\n (a) as already mentioned, proposed perturbation schemes may lead to results
incompatible with a probabilistic interpretation, 
particularly, violation of positivity of the states$^9$ 

\n (b) derivations based on the theory of stochastic
processes {\em a priori} impose conditions on the physical
characteristics of the systems studied (e.g. postulate a
master evolution equation for the states, like the
Chapman-Kolmogorov equation, on which approximations are
subsequently made).

 Instead, our aim in this paper is to
provide a mathematically rigorous and physically adequate
framework for the study of such kinetic equations, and examine
certain of the properties of their solutions, formally
determined by one parameter semigroups ~$V_t = {\rm e}^{Z_S
t}$.  Specifically, we give {\em minimal} conditions under which

\begin{itemize}
\item[(1)] Eq.~(1.1) {\em does} indeed define rigorously a
semigroup of solutions capable of a probabilistic
interpretation, particularly preserving positivity, 

\item[(2)] it is
possible to characterize $Z_S$ and $Z_O$ and explicit their
general form if they are differential operators, 

\item[(3)] an H-theorem for arbitrary continuous convex functionals of the
state can be proved.
\end{itemize}

\n In this way, the outline of the formal framework
for a probabilistic description of finite classical systems
given at the beginning of this section, is made precise and
kinetic equations like (1.1), (1.1$'$) become in
principle mathematically and physically meaningful.  Our
approach is similar to that followed in defining dynamical
semigroups for quantum systems$^{11}$. To this end,
we notice that (1.2) is essentially a duality relation,
suggesting that states are linear functionals defined on
observables.  Then the interpretation of states as probability
densities in (1.1) implies that observables are {\em bounded}
functions in  phase-space.  Then, in view of (1.2), finite
expectation values for {\em all} observables naturally leads to an
extension of the state-space beyond probability densities, so
that all linear functionals defined on bounded functions are
included.  In section 2, these qualitative remarks are
precisely formulated, thus establishing a sufficiently wide
mathematical framework for the derivation of rigorous
results.  In particular it will be seen that {\em kinetic equations
like (1.1$'$) should necessarily define a Markov semigroup of
solutions on the (Banach) space of observables}. In section 3
we collect basic theorems concerning the generator of a Markov
semigroup.  Some proofs are presented, mainly because they are
partly used as intermediate steps toward other
results so that the paper is sufficiently self-contained.
In section 4 we show rigorously that the only
differential operators that are candidates as generators of
Markov semigroups, are at most of the 2nd order, namely
(degenerate in general) elliptic operators (Lemma 4.1).  We
further supply a complete proof of the fact that under a certain
 continuity assumption, well-known in
the theory of stochastic processes, eq(4.1), the generator of a
Markov semigroup is indeed a differential operator (theorem
(4.1)). The {\em converse problem} is considered in section 5, where
we show that under mild regularity conditions, in general
{\em degenerate} elliptic operators generate Markov semigroups
{\em globally} defined on the space of observables.  This
generalizes classical results on local solutions of
nondegenerate elliptic equations. Finally in section 6, {\em a
general H-theorem for arbitrary continuous functionals on the
state-space is formulated and proved for any} (and not only
for a diffusion type) {\em Markov semigroup, under the assumption
that an invariant measure exists, which for a diffusion-type semigroup
is implied by the existence of 
an equilibrium solution $\rho_0$ of (1.1)} (theorem 6.1). It
should be emphasized at this point that $\rho_0$ {\em need not be
a probability density},  i.e.  integrable on the whole
phase-space of the system (e.g. for particle systems, in
position as well as in velocity space). This is a {\em global}
formulation of the H-theorem (corollary (6.1)) to be
contrasted to the {\em local} one given as an entropy continuity
equation, and usually obtained in cases the kinetic equation
for the states has a nonintegrable stationary solution.
Moreover, it will become apparent that {\em the existence of such
a solution} (satisfied by the formalism of Refs.6, 7)
implies that the Markov semigroup property (specifically,
positivity conservation) is (essentially) {\em equivalent to the
existence of H-functions}. 
\vskip 0.9truecm

 \begin{center}
{\bf 2. MATHEMATICAL ASSUMPTIONS AND THEIR\\ PHYSICAL MOTIVATION:
MARKOV SEMIGROUPS}
\end{center}
\bigskip
As already mentioned, inherent to
kinetic equations like (1.1) is a probabilistic
interpretation. Motivated by the preceding qualitative discussion,
we introduce a  {\em minimal} set  of assumptions to fix an
appropriate mathematical framework for states and observables
of a finite classical  system and for their time
evolution given by kinetic equations like (1.1), (1.1$'$).
Though the interpretation of states is better suited to
supply the {\em physical} motivation for such assumptions, it is
{\em mathematically} more convenient to work with 
observables. This is  suggested by theorem 2.1 below and
will become more evident in sections 3-5. The results thus
obtained will enable us to return to the state space when
formulating and proving the H-theorem in the last section. 

We consider admissible observable quantities $A$ as
(real) bounded continuous functions on the phase
space $X$, that remain finite at infinity, $X$ being a
differentiable manifold (e.g. the phase-space of a
Hamiltonian system). It is convenient to embed them in a
complex linear space (and extend $w$ in (1.2) to a sesquilinear
form). Thus we introduce
\smallskip

\n {\bf Assumption ($A_1$) }: The phase-space $X$ of a system
is a differentiable manifold, and observables belong to
$C(X_\infty,{\Cx})$, the space of all continuous complex
valued functions having a finite limit at infinity.

\smallskip
\n Here, $X_\infty$ is $X$ together with the infinite
point (more precisely, the one-point compactification of $X$).
Then  continuous functions $A$, having a finite limit at infinity
(in particular constant functions)$^{12}$
can be identified with continuous functions on $X_\infty$, by
 defining~$A(\infty)$ to be this finite limit.  Thus observables belong to
$C(X_\infty,{\Cx})$ which is a Banach space 
with norm $||A||={\rm sup}_{x\in X_\infty} |A(x)|$.

In section 1, it has been
suggested that states~$l$ are linear functionals on the space
of observables, their values~$l(A)$ giving expectations.
Then, their probabilistic interpretation requires them to be
positive, i.e. $l(A)\geq 0$ for $A\geq 0$, since expectation
values of nonegative random variables are nonegative (notice
that $l$ is real in this case). Then $l$
is bounded with bound~$||l||=l(1)$ $^{13}$, hence $l$ is continuous on
$C(X_\infty,{\Cx})$ that is $l\in C^\ast(X_\infty,{\Cx})$ where
$S^\ast$ is the Banach dual of a normed space $S$.  The above
comments are summarized in
\smallskip

\n {\bf Assumption ($A_2$) }: If $C(X_\infty,{\Cx})$
contains the observables $A$, then states $l$ are positive
linear functionals on it and expectation values are given by
$l(A)$.

\smallskip 
\n We remark here, that states thus
introduced, extend the class of states defined by probability
densities, this extended class being
characterized as probability measures by the Riesz-Markov
theorem below $^{13, 14}$.

\smallskip
\n {\bf Theorem 2.1} (i) For every positive 
$\ell \in C^\ast (X_\infty,{\Cx})$ there exists a unique {\em positive}
 (regular) Borel measure $\mu$ such that
$$
\ell(A)=\int\, A(x)d\mu (x) \ , \hskip0.5truecm 
\|\ell\|=\ell(1)=\mu (X_\infty) \eqno(2.1)
$$
\n (ii) More generally, if $\ell\in C^\ast (X_\infty,{\Cx})$,
there exists a unique {\em complex} Borel measure $\mu$ such
that (2.1) holds and
$\| \ell\|=|\mu| (X_\infty)$, where $|\mu|$ is the
total variation of $\mu$ $^{14}$.

\smallskip
\n {\em Remarks:} In the present work all measures are {\em regular
Borel measures}. The theorem is valid if $C(X_\infty,{\Cx})$ is replaced
by $C_0(X,{\Cx})$ or $C_c(X,{\Cx})$, the spaces
of functions vanishing at infinity or having compact support
respectively. Finally, we notice that unbounded phase space
functions such as polynomials, have been excluded from
observables, since for such functions it is impossible to have
finite expectation values for {\em all} states.

Having established states and observables as elements of
dual Banach spaces, we introduce for their time evolution (1.1), (1.1$'$)

\smallskip
 \n {\bf Assumption $(A_3)$:} Equation (1.1$'$) has a
well-posed  initial-value problem, that is:
for given initial data, the solution is unique for $t > 0$,
it depends continuously on the initial
data (small changes in the initial data imply small
changes in the solution, in the sense of the Banach space norm) and
expectations are continuous in $t$.

\smallskip
\n
Notice that $A_3$, proposition 2.2 below and $A_2$ imply that $A_3$ holds
for (1.1) as well and by the linearity of (1.1$'$)
 nonuniqueness for one initial
condition implies nonuniqueness for all. It is well known that $A_3$
implies 
\smallskip

\n {\bf Proposition 2.1} If (1.1$'$) is linear, autonomous and
has a well-posed initial value problem on some set $U$ of the
Banach space of observables, then its solutions $A(t)$ with initial
conditions $A$, uniquely define a s-continuous one-parameter semigroup of
operators $T_tA \equiv A(t)$ on the Banach subspace $B\equiv {\bar U}$ 
(from now on the prefix s or w means  strong or weak respectively and
$\bar U$ is the s-closure of $U$).  
\smallskip

\n Here we used  that continuity of expectation values imply
w-continuity of $T_t$ for $t=0$, which is equivalent to its
$s$-continuity for $t \geq 0$ $^{15}$ and
\smallskip

\n {\bf Definition 2.1} A (one-parameter s-continuous) semigroup (of
operators) on a Banach space $B$, is a family of bounded linear
operators  $\{$ $T_t,\ t\geq 0$ $\}$, such that:

(i) $T_{t+s}=T_tT_s$, \ 
(ii) $ \slim_{t\to t_0}T_tA=T_{t_0}A \qquad \forall A\in B, $
\smallskip

\n Its generator $Z$ is defined by 
$${\cal D}(Z)=\{ A\in B\ :\ 
\slim_{t \to 0^+}{T_tA-A\over t}\quad \mbox{exists} \} \ ,  \hskip
0.5truecm  Z A=\slim_{t\to 0^+}{T_tA-A\over t} \eqno(2.2)$$
It is well-known that the {\em converse} holds$^{16, 17}$ -
 derivatives and  integrals
being defined using the Banach space norm $^{16}$. 
\smallskip

\n {\bf Theorem 2.2} Let $T_t$ be a semigroup on a Banach 
space, $s$-continuous at $t=0$ on a Banach subspace $B$.
Define $Z$ by (2.2) $({\cal D}(Z) \neq \emptyset)$. Then

\n
(a) on $B$, $T_t$ is uniquely defined by $Z$ and is s-continuous for $t
\geq 0$;  $B$ is $T_t$-invariant, norm-closed and
$$
T_0|_B=I,\qquad B=\overline{{\cal D}(Z)}, \quad  \; \
\| T_t\| \leq Me^{at}\quad \mbox{for some}\quad M\geq 1,\quad
a\in {\Rl} \eqno(2.3)
$$
\n (b) $Z$ is a closed operator, for $A\in {\cal D}(Z)$, $T_tA$ is
$s$-differentiable (cf.(2.2)), hence $T_t({\cal
D}(Z))\subseteq {\cal D}(Z)$ 
$$
 {\partial T_t A \over \partial t}=Z T_t A=T_t Z A \quad \quad
T_t A-A=\int_0^t\,T_s A\,ds \eqno(2.4)
$$
\n and $T_tA$ is the unique solution of (2.4)
(cf.(1.1$'$)) on $B$ with initial condition $A$, 
satisfying $\| A_t\| \leq ce^{kt}$  for $c,\ k$ constants.
 
\n (c) For $\lambda  > a$ (cf.(2.3)) $\lambda -Z:
{\cal D}(Z)\rightarrow B$ is invertible, its inverse $R_\lambda $, the
resolvent of $Z$, is bounded and
$$
  R_\lambda A= \int_0^\infty e^{-\lambda t}T_t A\,dt \ ,
 \hskip 0.5truecm \| R_\lambda \| \leq {M\over {\lambda - a }} \eqno{(2.5)}
$$

\smallskip
The adjoint semigroup $T_t^\ast$ is everywhere defined on $B^\ast$ by
$(T_t^\ast \ell)(A)=\ell(T_tA)$. 
 Similarly since $Z$ is  densely defined on $B$, its
adjoint $Z^\ast$ is defined for all 
$\ell\in B^\ast$ for which $A\rightarrow \ell(ZA)$ is bounded,
as the unique  bounded extension of this functional to $B$. Thus
$  (Z^\ast \ell)(A)=\ell(ZA)\; \forall A\in {\cal D}(Z)$.
Now, unless $B$ is a reflexive Banach space neither
$T_t^\ast$ is  $s$-continuous on $B^\ast$ nor $Z^\ast$ is its
generator. Therefore (1.1$'$), (1.1) are not
simply adjoint equations. However we have $^{17}$
\smallskip

\n {\bf Proposition 2.2} Let $T_t$ be a $s$-continuous
semigroup on the Banach space $B$, with generator $Z$. Then
its adjoint $T_t^\ast$ is $s$-continuous on a
(presumably proper) Banach subspace of $B^\ast$, uniquely 
determined on $B^\ast$ by its generator however, and 
 for which $Z^\ast$ is an extension.

\smallskip
\n Thus going back to (1.1), (1.1$'$), either $Z_S$ or $Z^\ast_O$ can be
taken as  the generator in the evolution equation for the
states without  any inconsistency. 
In connection with $A_2$ we require conservation of positivity and of
normalization of the states for all times, that are indispensable for
their probabilistic interpretation
\smallskip

\n {\bf Assumption $(A_4)$:} 
(a) For $t > 0$,  $\ell$ positive implies $T_t^\ast \ell$ positive;
(b) $T_t^\ast \ell(1) = \ell(1)$ for any state $\ell$.
\smallskip

 \n By $A_4$ and theorem 2.1 
$$ 
T_t A \geq 0 \quad \mbox {for continuous}\quad A \geq 0 \quad
\mbox{in} \quad B ,  \quad \quad  T_t 1 = 1 \eqno(2.6) 
$$
\n which imply$^{17}$ that $\| T_t \| = \| T_t(1) \| = 1$, i.e.
$T_t$ (hence $T_t^\ast$)  is a {\em contraction semigroup}
($\| T_t \| \leq 1$).

In summary, assumptions $A_1$-$A_4$, sufficiently well
justified upon physical and mathematical considerations, imply that
linear autonomous kinetic equations conserving the probabilistic
interpretation of states, 
 {\em necessarily generate Markov semigroups on $C(X_\infty,{\Cx})$,
the space of observables, i.e. a positivity and normalization
preserving $s$-continuous semigroup} (cf.(2.6)).
 In the next two sections their properties are considered 
(proof of their 1-1 correspondence with
time-homogeneous Markov processes is outlined in Appendix 1).
\vskip 0.9truecm


\begin{center}
 {\bf 3. GENERAL RESULTS ON THE GENERATOR\\ OF A MARKOV
SEMIGROUP}
\end{center}
\bigskip
Here we summarize results on the generator of a Markov semigroup.
Though we follow mainly Ref.17, some proofs are
slightly modified so that a coherent presentation results.

\n The Hille-Yosida theorem is in our case $^{15, 17}$
\smallskip

\n {\bf Proposition 3.1} A closed, densely defined operator
$Z$ on a  (presumably complex) Banach space $B$ is the
generator of a $s$-continuous  contraction semigroup 
of operators iff its resolvent $R_\lambda =(\lambda
-Z)^{-1}$ is  an everywhere defined bounded operator for
$\lambda >0$ and
$$ 
\| R_\lambda \| \leq 1/\lambda ,  \qquad \lambda >0 \eqno(3.1)
$$
\n {\em Remark:} The resolvent set of $Z$, i.e. those $z\in
{\Cx} $ such that $R_z$ is everywhere defined, contains ${\Rl}^+$,
and since $Z$ is closed, it is an open set, 
 in particular $\mu$ belongs there
 for $|\mu -\lambda | \leq {1\over \| R_\lambda
\|}$.  $R_\lambda $  is a closed (hence
bounded, by the closed-graph theorem) operator on $B$. 

\smallskip
 Contractivity of a semigroup is related to the
dissipativity of its generator: by
the Hahn-Banach theorem, in a  normed space $B$ and for 
every $A\in B$ there exists an (in general nonunique)
$\ell\in B^\ast$ such that$^{15}$ 
$$
 \| \ell\| =1\quad ,\ \ell(A)=\| A\| \quad \mbox {equivalently}
\quad \| \ell\| =\| A\| \qquad \ell(A)=\| A\|^2 \eqno(3.2)
$$
\n Thus we give:
\smallskip

\n {\bf Definition 3.1} An operator $Z$ on a normed space $B$
is dissipative if and only if for every $(\ell,A)\in B^\ast
\times B$ satisfying (3.2) (Re denoting the real part)
$$ 
\mbox{Re}\: \ell(ZA)\leq 0  \eqno(3.3)
$$
\n {\em Remark:} For a Hilbert space with scalar product $\langle,\rangle,$
(3.3) is equivalent to 
$\mbox{Re}\, \langle A,ZA\rangle \leq 0$, hence 
$\| T_tA\|^2$ decreases monotonically in time (cf.(2.4)). This
is closely  related to an $H$-theorem for the adjoint equation
(section 6).
\smallskip

\n {\bf Proposition 3.2} If $Z$ is a densely defined {\em closed}
operator on a Banach space $B$, the following are equivalent:

\n (a) $Z$ generates a s-continuous contraction semigroup on
$B$.

\n (b) (i) $\mbox{Range}(\lambda -Z)=B\quad \forall \lambda
>0$, (ii) $Z$ is dissipative

\n (c) (i) $\exists \lambda >0 : \mbox{Range}(\lambda -Z)=B$, 
(ii) $\forall A\in B\; \exists \ell\in B^\ast$ so that (3.2),
(3.3) hold.
\smallskip

\n {\bf Proof:} (a)${\Rightarrow}$(b): Condition (b,i) is part
of proposition 3.1 Take $\ell$, $A$ as in (3.2). Then by the
continuity of $\ell$ and (3.2)
$$
\mbox{Re}\: \ell(ZA) =
\mbox{Re}\,\lim_{t\rightarrow 0^+}{\ell(T_tA)-\|A\|\over t}
\leq \mbox{Re}\, \lim_{t\rightarrow 0^+}{(\|T_t\|-1)\|A\|\over t}\leq 0
$$ 
\n (b)$\Rightarrow $(c) trivial

\n (c)$\Rightarrow$(a) for $A \in {\cal D}(Z)$ take $\ell$ as
in (3.2). Then for {\em any } $\mu  > 0$
$$ 
\| (\mu -Z)(A)\| \geq |\ell ((\mu -Z)A)| = |\mu\| A | -
\ell(ZA)|\geq\mu \| A |\
$$
\n hence $\mu -Z$ is 1-1 in its range, and if this
is equal to $B$, then $\mu$ is
in the resolvent set of $Z$ and $\| R_\mu\|\leq{1\over \mu }$; in particular  
 ${1 \over \| R_\lambda \|} \geq \lambda$.
 By the remark in proposition 3.1, every $\mu \in
 (0, {3\lambda \over 2}]  \subseteq (0, \lambda
+ {1 \over \| R_\lambda \|})$ is in the resolvent set of $Z$.
By induction all $\mu \in (0 ,({3\over 2})^n \lambda ]$
are in the resolvent set, for all $n \in {\Nl}$. \hfill  Q.E.D.
\smallskip

\n Modifying slightly the proof (c)$\Rightarrow$(a) above we obtain
$^{17}$
\smallskip

\n {\bf Corollary 3.1} If $Z$ is a densely defined {\em closable}
operator on the Banach space $B$ with
$\mbox{Range}(\lambda -Z)$ dense in $B$ for {\em some}
$\lambda > 0$ and 
condition (c,ii) of proposition 3.2 holds, then its closure
$\tilde{Z}$ generates
a $s$-continuous semigroup on $B$.
\smallskip

\n {\em Remark:} From (2.10-11)  conservation of {\em positivity and
 normalization} on $C(X_\infty, {\Cx}$), {\em imply dissipativity}.
 For a {\em partial} converse we have
\smallskip

\n {\bf Proposition 3.3} A $s$-continuous contraction semigroup
$T_t$ of operators on $C(X_\infty, {\Cx})$ with generator $Z$,
is positivity-preserving if  $(ZA)^\ast = ZA^\ast$, and
$$ 
\mbox{if}\quad A(x_0)\geq A(x)\; \forall x\in X_\infty\quad
\mbox{then}
\quad ZA(x_0)\leq 0  \eqno(3.4) 
$$
\n {\bf Proof:} By proposition 3.1, every $\lambda  > 0$ is in
the resolvent set of $Z$ and since ${\cal D}(R_\lambda ) =
C(X_\infty,{\Cx})$, we shall show that $(\lambda  - Z)A \geq
0\;\Rightarrow\; A \geq$ 0. Suppose $A$ takes negative
values on $X_\infty$, with an  absolute minimum	$A(x_0)$. 
By (3.4) $(\lambda  - Z)A(x_0) < 0$ for any $\lambda  > 0$
and by proposition 3.1, for any $\lambda>0$
$\| \lambda (\lambda R_\lambda  - I) \| {\leq} 2\lambda \;
 \forall \lambda  > 0$.
Therefore $T_t^\lambda \equiv e^{-\lambda t} e^{\lambda ^2 R_\lambda t} $
is a well-defined, positivity-preserving $s$-continuous
semigroup and $\lambda (\lambda  R_\lambda  - I)= \lambda  Z R_\lambda$.
From this follows that $\slim_{\lambda  \rightarrow +\infty}\lambda (\lambda 
R_\lambda  - I)A = ZA $
for $A\in {\cal D}(Z)$, which implies 
$ \slim_{\lambda  \rightarrow +\infty}T_t^\lambda  = T_t A $
uniformly in $t$ on bounded intervals$^{17}$.
Hence $T_tA \geq 0$ for $A \geq 0$ \hfill Q.E.D.
\smallskip

\n {\em Remark:} If $1\in {\cal D}(Z)$, (3.4) implies
$Z(1) = 0$, hence  $T_t1 = 1$.

\smallskip
\n
Proposition 3.2 and theorem A.1.1 give the characterization of
the generator  of a Markov semigroup by (see Ref.34, section
7.7 for a slightly different  formulation)
\smallskip

\n {\bf Theorem 3.1} Let $Z$ be a densely defined closed
operator on $C(X_\infty, {\Cx})$. Then the following are
equivalent:

\n (a) $Z$ generates a Markov semigroup $T_t$

\begin{tabbing}
\n (b) \=(i)  $1 \in {\cal D}(Z)$,\\
	 \>(ii) $\exists\lambda > 0:\mbox{Range}(\lambda-Z)=
	 C(X_\infty,{\Cx})$,\\
	 \>(iii) $(ZA)^\ast = ZA^\ast\;\forall A \in {\cal D}(Z)$,\\
	 \>(iv) (3.4) holds.
\end{tabbing}

\n {\bf Proof:}  (a) ${\Rightarrow}$ (b). (i) follows from
(2.6), (ii) from proposition 3.1 since $T_t$ is a contraction
semigroup. By theorem  A.1.1 and (A.1.2) both (iii), (iv) are evident.

\n (b) ${\Rightarrow}$ (a). Let $A\in{\cal D}(Z)$. Since
$X_\infty$ is compact, $|A|$ continuous,  $\|A\| = |A(x_{0})|$
for some $x_0 \in X_\infty$. Without loss of generality let $A(x_0)>0$ and 
$\delta _{x_0}$ be the $\delta$-function at $x_0$, which is a
positive  linear functional on $C(X_\infty,{\Cx})$,
hence $\|\delta _{x_0}\| = \delta _{x_0}(1) = 1$. Then
$\|A\| = A(x_0) = \delta _{x_0}(A) \geq A(x)\; \forall x \in
X_\infty$, so that by (3.4) $\delta _{x_0}(Z A) = Z A(x_0) \leq
0 $. Hence  by (ii) and proposition 3.2, $Z$ generates a
contraction semigroup, which by (ii), (iii) and
proposition 3.3 is positivity-preserving.  Since ${\cal
D}(Z)$ $\ni 1,\; T_t1=1$ by the remark to proposition 3.3.
\hfill Q.E.D.
\smallskip

\n {\em Remark:} 
Proposition 3.3 and theorem 3.1 show that {\em (3.4)  is the
crucial condition relating dissipativity with conservation of
positivity} (see end of section 4 and section 6).
 \vfill{\eject}


\begin{center}
 {\bf 4. DIFFERENTIAL OPERATORS GENERATING\\ MARKOV SEMIGROUPS} 
\end{center}
\bigskip
In view of the last remark
we explicit the form of the generator of a Markov semigroup that is a
{\em differential} operator, since kinetic equations are often given
by such operators. We first prove$^{18, 19}$
\smallskip

\n {\bf Lemma 4.1} Let $Z$ be a differential operator of order
$k$, defined on $C^k({\Rl}^n,{\Rl})$ and suppose that for  any
$f \in {\cal D}(Z)$, if $f(\vec{x}_0) \geq f(\vec{x})\;
\forall \vec{x} \in {\Rl}^n$ then $Z f(\vec{x}_0) \leq 0$.
Then $Z$ is at most a 2nd order differential operator with
nonegative-definite 2nd order coefficient.

\n {\bf Proof:} (i) Let $n = 1$.  By the remark to proposition 3.3, $Z$ has
no zeroth order term. Suppose $Z$ has the form
$$ 
Z = c_1(x) {d \over dx}+c_2(x) {d^2 \over dx^2} + c_k(x)
{d^k \over dx^k }\quad k \geq 3 
$$
and that $c_k(x_0) \neq 0$ for some $x_0\in {\Rl}$. Let
$$ 
g(x) = -\epsilon  (x - x_0 )^2 + a (x - x_0 )^k ,\quad \epsilon  > 0,\;
a {\in} {\Rl} 
$$
 Clearly $g$ has a local maximum zero at $x_0$. There is a
neighborhood $N(x_0)$  with compact closure, in which $g(x)
\leq g(x_0) = 0$.  Let $\tilde{f}(x)$ be a
$C^k$-function with compact support, $\tilde{f}(x) \leq 0$ on
${\Rl} - N(x_0)$ having the same partial derivatives with $g$
up to order $k$, on the boundary of $N(x_0)$. Then 
$$
f(x) = \cases {g(x) &, $x \in N(x_0)$ \cr \tilde{f}(x) &, $x
\in {\Rl} - N(x_0)$ \cr } 
$$
is a $C^k$-function with $f(x)\leq f(x_0) = 0\;\forall x
\in {\Rl}$ and $Zf(x_0) = -2\epsilon c_2(x_0) + k! a c_k(x_0)$.
Since $k>2$, appropriate choice of $a$, makes it positive, unless
$c_k(x_0)=0$.

\n
By a similar argument $c_2(x)\geq0\;\forall x\in{\Rl}$ 
(for any $x_0 \in {\Rl}$ take $g(x) = -\epsilon (x - x_0)^2, \epsilon > 0$).

\n (ii) For $n > 1$ let $Z$ contain a term of the form 
$c(\vec{x})\partial_{i_1}^{a_1}\partial_{i_2}^{a_2}...
\partial_{i_\ell}^{a_\ell}$, where
$a_1 + a_2 +...+ a_\ell = k > 2 $ and $\partial_i = {\partial
\over \partial x^i }$. The proof can be repeated,
starting with a $\vec{x_0}$ for which $c(\vec{x_0})\neq0$ and
$$ 
g(\vec{x}) = a(x^{i_1} - x_0^{i_1})^{a_1}...(x^{i_l} - 
x_0^{i_\ell})^{a_\ell}-\epsilon  \sum_{i=1}^n (x^i - x_0^i)^2 
$$
since it has a local maximum at $\vec{x}_0$. With
$g(\vec{x}) = -\sum_{i,j=1}^n \lambda _i
\lambda _j (x^i - x_0^i) (x^j - x_0^j),\ \lambda_i \in {\Rl}$,
 proceeding as before we obtain
$ Zf(\vec{x}_0) = -\sum_{i,j=1}^n a_{ij}(\vec{x}_0)\lambda_i
\lambda_j \leq 0 $
where $a_{ij}$ is the 2nd order coefficient of $Z$. \hfill Q.E.D.

\n {\bf Remark}: The proof above is local. Therefore working  in some local
coordinates of a differentiable manifold and using its local  compactness,
so that $N(\vec{x_0})$, hence $\tilde f$ above exist, the
same proof holds.

\smallskip
\n Now $C_c^2({\Rl}^n) \subseteq C({\Rl}_\infty^n)$,  with the
identification
of continuous functions of ${\Rl}^n$ having a finite limit at
$\infty$ with elements of $C({\Rl}_\infty^n)$ made in section
2. Then the following theorem supplies a 
{\em sufficient} condition for the generator of a markov semigroup to
be a differential  operator$^{20}$.
\smallskip

\n {\bf Theorem 4.1} Let $T_t$ be a Markov semigroup on
$C({\Rl}_\infty^n,{\Cx})$ with generator $Z$ and suppose that

\n (i) $C_c^2({\Rl}^n) \subseteq {\cal D}(Z)$

\n (ii) The transition probability measure $p(t,\vec{x},
\cdot)$ of the corresponding Markov process (see appendix 1) satisfies
$$ 
\lim_{t \rightarrow 0^+} {p(t, \vec{x}, N) \over t} =
\chi_N(\vec{x})\quad \mbox{uniformily in}\quad \vec{x}
\eqno(4.1) 
$$
Then
$$
Zf(\vec{x}) =  b_i(\vec{x}) \partial_i
f(\vec{x}) + a_{ij}(\vec{x}) \partial_i\partial_j
f(\vec{x}) \quad f \in C^2({\Rl}^n) \eqno(4.2) 
$$
$b_i,a_{ij}$ are continuous, real-valued functions, $a_{ij}$
is a nonegative-definite matrix $\forall \vec{x} \in {\Rl}^n$,
$\chi_N$ is the characteristic function of $N$ and the summation
convention is henceforth assumed.
\smallskip

\n {\bf Proof:} Let $\vec{x}_0 \in {\Rl}^n$,  and for $f \in
C_c^2({\Rl}^n)$ consider its 2nd order Taylor polynomial at
$\vec{x_0}$, denoting by $\tilde{f}(\vec{x})$ the corresponding
remainder. By Taylor's theorem, for every $\epsilon > 0$,
$\tilde{f}(\vec{x})\leq \epsilon \|\vec{x}- \vec{x}_0\|^2$
in some neighborhood $N(\vec{x}_0)$. Put
$$ 
g_\epsilon (\vec{x}) =  \cases { \epsilon 
\|\vec{x}-\vec{x}_0\|^2  &, $\vec{x} \in N(\vec{x}_0)$ \cr 
\sigma(\vec{x}) &, $\vec{x} \in {\Rl}_\infty^n -
N(\vec{x}_0)$ \cr } 
$$
where $\sigma(\vec{x})\in C_c^2({\Rl}^n)$ is nonnegative and
it and its first two partial derivatives coincide with those
of $\tilde{f}$ on $\partial N(\vec{x}_0)$ (without loss of
generality we take $\overline{N(\vec{x}_0)}$ compact). 
Then $|\tilde{f}(\vec{x})|\leq \epsilon \ g_\epsilon (\vec{x}),
\quad \vec{x} \in N(\vec{x}_0)$ and
$$ 
| Z\tilde{f}(x_0)| = \lim_{t \rightarrow 0^+} | {1 \over t}
\int_{{\Rl}_\infty^n-N(\vec{x}_0)}\tilde{f}(\vec{y})
p(t,\vec{x}_0,d\vec{y}) +{1 \over
t}\int_{N(\vec{x}_0)}\tilde{f} (\vec{y}) p(t,
\vec{x}_0,d\vec{y})| 
$$
By (4.1) $ |{1 \over t} \int_{{\Rl}_\infty^n-N(\vec{x}_0)}
\tilde{f}(\vec{y})p(t,\vec{x}_0,d\vec{y})| \leq \|\tilde{f}\| o(1)$.
\n On the other hand, since $g_\epsilon \geq 0 $  
${1 \over t} \int_{N(\vec{x}_0)}\tilde{f}(\vec{y}) p(t,\vec{x}_0,d\vec{y})
\; \leq {\epsilon \over t} \int g_\epsilon(\vec{y}) p(t,\vec{x}_0, d\vec{y})$. 
Therefore $|Z\tilde{f}(\vec{x}_0) | \leq\epsilon Zg_\epsilon
(\vec{x}_0)$.
But $\epsilon$ is arbitrary, hence $Z\tilde{f}(\vec{x}_0) = 0$.

\n Considering the functions $f^i(\vec{x}) = x^i - x_0^i$, \
$f^{ij}(\vec{x}) \equiv (x^i - x_0^i)(x^j-x_0^j)$
in $N(\vec{x}_0)$, extended to $C_c^2({\Rl}^n)$-functions and using
(4.1) we get
$$ 
Zf^i(\vec{x}) =\lim_{t \rightarrow 0^+} {1 \over t}
\int_{N(\vec{x}_0)} p(t, \vec{x}, d\vec{y}) (y^i - x^i) 
$$
$$ 
Zf^{ij} =\lim_{t \rightarrow 0^+} {1 \over t}
\int_{N(\vec{x}_0)} p(t, \vec{x}, d\vec{y}) (y^i - x^i) (y^j -
x^j) 
$$
\n Returning to $f$ and using this and that $Z(1)=0$, we obtain
  $Z\tilde{f}(\vec{x}_0) = 0$. Since $\vec{x}_0$
is arbitrary, (4.2) follows with $b_i(\vec{x}) =
Zf^i(\vec{x}),\; a_{ij} = {1 \over 2} Zf^{ij}(\vec{x})$. By
the definition of $Z$, $b_i, a_{ij}$ are continuous and evidently
$a_{ij}(\vec{x})$ is nonegative-definite. \hfill  Q.E.D.
\smallskip

\n {\em Remarks:}\\ 
(a) Eq(4.1) implies the existence of the 
first two moments of $p(t, \vec{x}, \cdot)$ and
Lemma 4.1  that	all higher moments vanish$^{21}$.
Notice that in the theory  of
stochastic processes, (4.2) is usually derived by assuming
(4.1)  {\em and the existence} of the first two moments of
$p$, from which the  vanishing of higher moments follows$^5$.
Notice also that (4.1) implies uniform
stochastic continuity of $p(t, \vec{x}, N)$, (A.1.3).

\n (b) For arbitrary differentiable manifolds, theorem 4.2 holds,
since only local compactness of ${\Rl}^n$ and Taylor's theorem are used.

\n (c) The theorem is true if ${\Rl}^n ,{\Rl}_\infty^n$ are
replaced by $U,\;\overline{U}$, where $U\in{\Rl}^n$ is a
relatively compact  open  set, i.e. $\overline{U}$ is compact.

\smallskip
Theorem 4.1 shows that (4.1) is a rather strong condition. 
It should be desirable to have a characterization of the
generator of a Markov semigroup relaxing (4.1), namely
operators satisfying (3.4)$^{22}$.
In this connection the following remarks are relevant:
it can be shown that a {\em necessary and
sufficient} condition for a linear operator acting on $C^2$-
functions and for which $Z(1)=0$, is a 2nd order differential
operator, is that for any $A$ in its domain and $x_0$ such
that $A(x_0)=0$, it follows that $Z(A^3)(x_0) = 0$ \ $^{23}$.
On the other hand, (3.4) implies that for all
nonpositive functions $Z(A^3)(x_0) \leq 0$ whenever
$A(x_0)=0$.  Therefore in theorem
4.1 the equality-sign case is characterized.  To put it
differently : generators of Markov semigroups that are not only
differential operators, are characterized by the condition
that there exist nonpositive functions $A$ for which
$$
Z(A^3)(x_0) <0 \hskip 15pt {\rm when} \hskip 15pt  A(x_0) = 0
$$
Thus the problem reduces to the characterization of such
operators. 

\vskip 0.9truecm


\begin{center}
{\bf 5. DEGENERATE ELLIPTIC EQUATIONS
 AND MARKOV SEMIGROUPS}
\end{center}
\bigskip
In the previous section we have shown that {\em the generator
of a Markov semigroup is a 2nd order differential operator
if it is defined for $C^2$-functions and satisfies (4.1)}.  On the other
hand, as mentioned in section 1, various approaches to kinetic
theory that starting from microscopic classical dynamics and
imposing more or less systematic approximation schemes, lead
to linear autonomous kinetic equations for particular classes
of systems that are partial differential equations
(see e.g.\ the approaches based on iteration schemes
applied to the so-called generalized master equation of one
or another form$^{2-4}$).
Therefore {\em they are incompatible with
$A_1-A_4$ unless they are at most of the 2nd order with
nonegative-definite leading coefficient}.

Using the results of section 4 we shall show that under quite general
conditions, {\em the converse is true, namely an operator of the
form (4.2) generates a Markov semigroup on
observables globally defined on the phase space}. 
Moreover conservation of positivity is essentially  equivalent
to the existence of an $H$-theorem for the 
adjoint semigroup. This is the subject of the next  section. 
We consider ${\Rl}^n$, but our results are applicable to any
differentiable manifold, without essential modifications,
since they depend on local differentiability properties of
function of ${\Rl}^n$ and its local compactness.

We define
$$
C^2({\Rl}^n_\infty,{\Cx})=\{ f\in C^2({\Rl}^n,{\Cx})\; :\; f,
\partial_i f,\partial_i\partial_j f 
 \; \mbox{have a finite limit at infinity} \} \eqno(5.1)
$$
(cf. the comments following $A_1$ in
section 2), hence $C^2({\Rl}_\infty^n,{\Cx})$ is
identified with a subset of $C({\Rl}_\infty^n,{\Cx})$; 
 $C^k({\Rl}_\infty^n,{\Cx})$ is defined  similarly.
These are dense subsets of $C({\Rl}_\infty^n,{\Cx})$, noticing
that for any $\vec{x}_0\in {\Rl}_\infty^n$, and
$$
f(\vec{x})\equiv \cases{\exp{(-\| \vec{x}-\vec{x}_0
\|^2)} &, $\vec{x}\in {\Rl}^n$ \cr 0 &, $\vec{x}=\infty$ \cr }\eqno(5.2)
$$
\n $f$ is in $C^k({\Rl}_\infty^n,{\Cx})$ and
$f(\vec{x}_0)\ne f(\vec{x})$ if $\vec{x}_0\ne \vec{x}$ and using
the Stone-Weierstrass theorem.
\smallskip
\n {\bf Definition 5.1} The operator
$Z:C^2({\Rl}_\infty^n,{\Cx}) \rightarrow
C({\Rl}_\infty^n,{\Cx})$ is defined by
$$
Zf(\vec{x})=a_{ij}(\vec{x})\partial_i\partial_j f(\vec{x})
+b_i(\vec{x})\partial_if(\vec{x}) \eqno(5.3)
$$
where $a_{ij},b_i\in C({\Rl}_\infty^n,{\Cx})$. Here and in
what follows the summation convention is used.
By the above discussion $Z$ is densely defined. We next prove
\smallskip

\n {\bf Proposition 5.1} $Z$ in (5.3) is closable.
\smallskip

\n {\bf Proof:} (i) It is sufficient that if $\lim_{n\to
+\infty}\| f_n
\| =0$ for $f_n\in {\cal D}(Z)$ and there exists $g$ such that
$\lim_{n\to +\infty}\| Zf_n-g\| =0$ then $g=0$ $^{15}$.
Let $a_{ij},b_i$ be in $C^2({\Rl}_\infty^n,{\Cx})$ and take 
$\rho_0\in C^2({\Rl}_\infty^n,{\Cx})$ with 
$$
\rho_0(\vec{x})>0,  \quad \lim_{\vec{x}\to
\infty}\rho_0(\vec{x})=\lim_{\vec{x}\to \infty}
\partial_i\rho _0(\vec{x})=0, \qquad   
\int_{{\Rl}^n}\rho _0(\vec{x})d\mu (\vec{x})=1 \eqno(5.4)
$$
where $\mu$ is the Lebesgue measure of ${\Rl}^n$. Clearly
$\rho_0$ has the form
$$
\rho_0(\vec{x})=c\exp(-\beta H(\vec{x}))\qquad c,\;\beta \in 
{\Rl}^+ \eqno(5.4')
$$
\n and
$$
\tilde{\mu} :\; \tilde{\mu} (E)\equiv \int_E
\rho_0(\vec{x})d\mu (\vec{x}) \eqno(5.4'')
$$
defines a measure on ${\Rl}^n$,  extended to a
regular measure on ${\Rl}^n_\infty$  by putting
$p(t,\vec{x}, \infty)=0$. By (5.4)and Lemma 5.1 below 
$$
{\cal D}(Z),\; \mbox{Range}(Z)\subseteq C({\Rl}_\infty^n,
{\Cx})\subseteq
L_{\tilde{\mu}}^\infty({\Rl}_\infty^n)\subseteq
L_{\tilde{\mu}}^2({\Rl}_\infty^n)\subseteq L_{\tilde{\mu}}^1
({\Rl}_\infty^n) \eqno(5.5)
$$
each $L_{\tilde{\mu}}^p$-space equipped with the usual
$\| \; \|_p$ norm, \ $p<+\infty$.

\n (ii) Let $\phi\in{\cal D}(Z)$ with compact support $\Omega$.
Using (5.4) and (5.4$'$), integration  by  parts gives
$$
\int\phi^\ast Zf_n d\tilde{\mu}=\int f_n[\partial_i\partial_j
(a_{ij}\rho_0\phi^\ast)-\partial_i(b_i\rho
_0\phi^\ast)]d\mu\equiv
\int f_n Z^\dagger(\rho_0\phi^\ast)d\mu \eqno(5.6)
$$
$$
Z^\dagger(\phi \rho_0)=\rho_0(Z\phi -2(\beta a_{ij}\partial_j
H-\partial_j a_{ij}+b_i)\partial_i\phi)+\phi Z^\dagger\rho_0\equiv 
\rho_0(Z\phi -H_i\partial_i\phi )+\phi Z^\dagger\rho_0 \eqno(5.7)
$$
\n $Z^\dagger$ being the formal adjoint of $Z$. Therefore, by (5.4) 
$$
|\int\phi^\ast g d\tilde{\mu}| \leq \|Zf_n-g\|\, \|\phi\|_1
+| \int\phi^\ast Zf_n d\tilde{\mu}|
\stackrel{n\to\infty}{\tini} 0
$$
\n Thus $\int\phi^\ast g d\tilde{\mu}=0$ for 
$\phi \in C_c^2({\Rl}^n, {\Cx})$  which is dense in $C_c({\Rl}^n, {\Cx})$.
Therefore the result follows  by the continuity of $\| \ \|_2$ and
Lemma 5.2 below.

\n (iii) For arbitrary $a_{ij},b_i$ in
$C({\Rl}_\infty^n,{\Cx})$ and $\epsilon >0$ there exist
functions $\hat{a}_{ij},\hat{b}_i\in
C^2({\Rl}_\infty^n,{\Cx})$ such that $\| \hat{a}_{ij}-a_{ij}\|
<\epsilon ,\ \| \hat{b}_i-b_i
\| <\epsilon $. Defining ${\hat Z}$ by (5.3) in terms of
  $\hat{a}_{ij},\hat{b}_i$ we get
$$
| \int\phi^\ast Zf_n d\tilde{\mu}|  
\leq \epsilon \| \phi \| (\|\partial_i\partial_j f_n\|
+\|\partial_if_n\|) +\| \phi \| \| {\hat Z}f_n\|\qquad \forall
\epsilon >0
$$
\n for any $\phi \in C_c^2({\Rl}_\infty^n,{\Cx})$ and we proceed
 as before. \hfill      Q.E.D.
\smallskip

\n {\bf Lemma 5.1} If $(X,$B$,\tilde{\mu})$ is a measure space
with $\tilde{\mu}(X)=1$ then for any $f$ in
$L_{\tilde\mu}^\infty(X)$ we have 
$\| f\|_p\leq \| f\|_q \quad \forall\; 0<p<q\leq \infty$
hence $ (L_{\tilde\mu}^q(X),\| \;\|_q)\subseteq (L_{\tilde\mu}^p(X),
\| \; \|_p)$  in a natural way$^{14}$
\smallskip

\n {\bf Lemma 5.2} If $f\in C({\Rl}^n,{\Cx}),\; \tilde{\mu}$
is a  finite measure on ${\Rl}^n$ and 
 for any $\phi \in C_c({\Rl}^n,{\Cx})$  $\int
f(\vec{x})\phi(\vec{x}) d\tilde{\mu} (\vec{x})=0$, then $f=0$.
\smallskip

Proposition 5.1 holds if
${\Rl}^n$, ${\Rl}_\infty^n$ are replaced by $U,\ \bar{U}$
respectively for any relatively compact open $U\subseteq
{\Rl}^n$. Obviously (5.4$'$) has to be modified.
Therefore, let $Z_{\bar U}$ be the
restriction of $Z$ in (5.3) on $\bar{U}$ for any relatively compact open 
$U\subseteq {\Rl}^n$, and $\tilde{Z}_{\bar U}$ its 
closure, easily seen to be the restriction
of $\tilde Z$  on $\bar U$. Then we have:
\smallskip

\n {\bf Proposition 5.2}  Suppose that for $Z$ in (5.3),
$a_{ij},b_i$ are in $C^2({\Rl}^n,{\Rl})$, $a_{ij}$ is a
nonegative - definite matrix and $\partial_k\partial_\ell a_{ij}$
are H\"older continuous for some  positive
exponent$^{24}$.
Then there is a (topological) base of relatively compact
(open) sets $\{U_a\}$ and a  $\lambda >0$ such that for each
$U_a$ $\mbox{Range}(\lambda -\tilde{Z}_a)=C(\bar{U}_a,{\Cx})$,
where $Z_a \equiv Z_{\bar{U}_a}$.
\smallskip

The proof$^{25}$ uses the
generalization of classical  results concerning the solution
of elliptic equations with {\em strictly}  positive-definite
2nd order coefficient, bounded from below away from zero$^{26}$, 
at the expense of requiring the coefficients to be
$C^2$-functions$^{27}$.
Relaxing the restriction of strict positivity of $(a_{ij})$ is
important for kinetic theory, given that it is violated by important
kinetic equations  (e.g. the (linear) Landau
equation either in phase-space or  velocity space, or model
equations like Kramers equation$^{2, 6, 7}$), whereas regularity
conditions are not so essential, since sufficient differentiability
is often implicitly required.

In the above notation, let
$f\in  C({\Rl}^n, {\Cx})$ and  $f_a\equiv f|_{\bar{U}_a}$.
Then there exists  a unique $u_a\in {\cal D}({\tilde
Z}_a)$ with  $(\lambda -{\tilde Z}_a) u_a=f_a$, 
hence  $\overline{\mbox{Range}(\lambda -Z_a)}=
C({\bar U}_a,{\Cx})$. Proceeding as in the 
proof of theorem 3.1, condition (cii) of proposition 3.2 is
 {\it equivalent} to the  nonegative-definiteness of 
$a_{ij}$ in (5.3). By Corollary 3.1, ${\tilde Z}_a$ is the
(dissipative) generator of a $s$-continuous contraction semigroup on
 $C({\bar U}_a,{\Cx})$.
Since  ${({\tilde Z}f)}^\ast$ $={\tilde Z}f^\ast$, and $1\in {\cal
D}(Z_a)$,  by theorem 4.1  ${\tilde Z}_a$ {\em	uniquely} defines a Markov
 semigroup on $C({\bar U}_a, {\Cx})$.
Restricting it on $C(U_a,{\Cx}),\ T_t^a$ say, we
define $(T_tf)(\vec{x})=(T_t^a f)(\vec{x})\quad \forall\; \vec{x}\in U_a $
for  $f\in C({\Rl}^n,{\Cx})$

By the uniqueness of $T_t^a, \ T_t$ is well-defined, positivity
preserving and since $U_a$ is open, $T_tf$ is continuous.
 Then for $f \in C_c({\Rl}^n,{\Cx})$, its support can be covered by a
 finite subset of $\{U_a\}$, $U_1,\cdots,U_n$ say, and  
$\mbox{support}(T_tf) \subseteq \mbox{support}(f)$, since on
each $U_a$, ${T_t}^a$ is a Markov semigroup. Therefore
$$
|T_t f(\vec{x})| \leq \max\{\| {T_t}^kf\|,\; k=1,2,\cdots,n\}\; \leq \| f\|
$$
hence $T_t$ is bounded. By the remark to theorem
2.1, any $\ell$ in $(C_c({\Rl}^n,{\Cx}))^\ast$ is
biuniquely defined by a  finite  measure $\mu$, so that
$$
\ell(T_tf-f)\leq \sum_{k=1}^n{\int_{U_k} d\mu (\vec{x})| T_t^k
f(\vec{x})-f(\vec{x})|}\leq \mu ({\Rl}^n)\sum_{k=1}^n\|
T_t^kf-f\|
$$
Thus $T_t$ is $w$-continuous at $t=0$, hence $s$-continuous
for all $t>0$. Therefore $T_t$ is uniquely extended to
$C_0({\Rl}^n, {\Cx})$ since $C_c({\Rl}^n, {\Cx})$  is dense
there$^{14}$, preserving the above properties (this extension is also denoted 
by $T_t$). As in the proof of theorem A.1.1, we show that $T_t$
satisfies (A.2.1) for some probability measure
$p(t,\vec{x},{\cdot})$ on ${\Rl}^n$, 
regularly extended on ${\Rl}_\infty^n$ by putting
 $p(t,\vec{x},\{\infty \})=0$. Thus $T_t$ is a Markov
semigroup on $C({\Rl}_\infty^n,{\Cx})$ with generator
locally given by the closure of $Z$ in (5.3). 
This completes the proof of 
\smallskip

\n {\bf Theorem 5.1} $Z$ in (5.3) is a densely defined,
closable operator on $C({\Rl}_\infty^n, {\Cx})$. Its closure
${\tilde Z}$ generates a Markov semigroup $T_t$ on
$C({\Rl}_\infty^n,{\Cx})$ if $(a_{ij})$ is a
nonegative-definite matrix function, $a_{ij},b_i$ are in 
$C^2({\Rl}^n, {\Rl})$ and $\partial_k\partial_\ell a_{ij}$ are
H\"older continuous for some exponent in (0,1).
\smallskip

\n {\em Remark:} Theorem 5.1 is valid
without essential modifications on any differentiable
manifold, in particular for a relatively compact open subset of
${\Rl}^n_\infty$ (see the remark preceding proposition 5.2).
\vskip1.2cm


\centerline{\bf 6. THE $H$-THEOREM AND CONCLUDING REMARKS}

\bigskip
 
In this section we consider arbitrary Markov semigroups on
$C({\Rl}_\infty^n,{\Cx})$, the generator of which {\em is not
necessarily} a differential operator.  We shall prove that {\em the
adjoint semigroup giving the evolution of states, satisfies a
general H-theorem}.  Working with the adjoint semigroup, the
proof turns out to be simpler and the theorem more general
than that obtained directly from the kinetic equation.
We introduce the measure $\tilde{\mu}$ defined by
(5.4$''$). Then any $\phi\in C({\Rl}_\infty^n,{\Cx})$ (or
$\phi\in L_{\tilde\mu}^1({\Rl}_\infty^n)\,$!) defines 
a bounded linear functional $\ell_\phi$ on $C({\Rl}_\infty^n,{\Cx})$ via
$$ 
\ell_\phi(A)=\int A(\vec{x}) \phi (\vec{x})
d\tilde{\mu}(\vec{x}) \eqno(6.1)
$$
If $T_t$ is a Markov semigroup then by theorem A.1.1 and
Fubini's theorem $^{14, 28}$, to $T_t^\ast \ell_\phi $
corresponds uniquely the measure (cf. (A.1.4)).
$$
\nu_t(E)\equiv \int p(t, \vec{y}, E)
\phi (\vec{y})d\tilde{\mu}(\vec{y}) \ , 
\hskip 0.5truecm (T_t^\ast \ell_\phi )(A) = \int A(\vec{x}) d\nu_t(\vec{x})
  \eqno(6.2)
$$
and we have the following
\smallskip

\n {\bf Theorem 6.1 ($H$-Theorem)} Let $T_t$ be a
Markov semigroup on $C({\Rl}_\infty^n,{\Cx})$, with corresponding
Markov process $p(t, \vec{x}, E)$. If
 $$ 
p(t,\vec{x},E)=\int_{E} q(t,\vec{x},\vec{y})d\mu (\vec{y}) \eqno(6.3)
$$
where $q(t,\cdot,\cdot)$ is jointly Borel measurable, and
$\tilde{\mu}$ is an invariant measure of $p(t,\vec{x},E)$, i.e.
$$
\int p(t, \vec{x}, E) d\tilde{\mu}(\vec{x}) =\tilde{\mu}(E) \eqno(6.4) 
$$
Then for any positive $\phi\in C({\Rl}_\infty^n,{\Cx})$,
$T_t^\ast \ell_\phi$ is defined uniquely by the density
$$
\phi_t(\vec{x}) = \int q(t, \vec{y},
\vec{x})\phi(\vec{y})d\tilde{\mu} (\vec{y}) \eqno(6.5)
$$
For any convex function $h: {\Rl}^+ \to {\Rl}$
and all nonegative $\phi\in C({\Rl}_\infty^n,{\Cx})$,
with $h\circ\phi $ $\tilde{\mu}$-integrable, 
$$
H(t)\equiv \int h\left({\phi_t\over\rho_0}\right)d\tilde{\mu} \eqno(6.6) 
$$
is a decreasing function ($H$-function) \ $^{29}$.
\smallskip

\n {\bf Proof:} (6.5) follows from (6.2), (6.3). For
the second assertion we use Jensens' inequality$^{14, 30}$.
Since $q(t,\cdot,\vec{x})$ is $\tilde{\mu}$-integrable,
$E\to\int_E q(t,\vec{y},\vec{x})d\tilde{\mu}(y)$ 
defines a measure on ${\Rl}^n,$  with a finite
total measure which by (6.3), (6.4) is $\int q(t, \vec{y},
\vec{x})  d\tilde{\mu}(\vec{y})	= \rho_{0}(\vec{x})$ $^{31}$.
Since $q,\;\phi$  are nonegative, Jensen's inequality and
(6.5) give
$$
h\left(\int {q(t,\vec{y},\vec{x})\phi(\vec{y})\over
\rho_0(\vec{x})} d\tilde{\mu}({\vec y)}\right)=h\left({\phi
_{t}(\vec{x}) \over
\rho_{0}(\vec{x})}\right)\leq\int{q(t,\vec{y},\vec{x})
h(\phi (\vec{y}))\over \rho_{0}(\vec{x})} d\tilde{\mu}(\vec{y}) 
$$
Therefore by $\int q(t,\vec{y},\vec{x}) d\mu(\vec{x}) =
p(t,\vec{y},{\Rl}_\infty^n)=1 $
$$ 
\int h\left({\phi_t(\vec{x}) \over \rho_{0}(\vec{x})}\right)
d\tilde{\mu}(\vec{x}) \leq \int h(\phi (\vec{y})) d\tilde{\mu}(\vec{y}) =
\int h \left({\phi_t(\vec{x})\over\rho_0(\vec{x})}\right)|_{t=0}
d\tilde{\mu}(\vec{x}) 
$$
and since $p(0, \vec{x}, {\vec E}) = \chi _{E}(\vec{x})$
by (6.3), $q(0, \vec{x}, \vec{y})= \delta (\vec{x}  - \vec{y})$
(cf. remark (b) to theorem A.1.1). Thus $H(t)\leq H(0)$.
If $\phi_t \equiv S_t\phi$ then by (6.5) and (A.1.1),
(6.3) implies the semigroup property for 
$S_t$, hence applying the above result to
$\phi_t$ for any $s \geq 0$ we have
$$ 
H(t+s) = \int h\left({S_s (S_{t}\phi ) \over \rho_0 }\right)
d\tilde{\mu} \leq \int h\left({S_t\phi\over\rho_0}\right)
d\tilde{\mu} = H(t)
$$
\hfill Q.E.D.

\n {\em Remarks:}\\ 
(a) Equation (6.4) is equivalent to $T_t^\ast \ell_1= \ell_1$
(c.f. (6.1))

\n (b) When $T_t$ is a
stochastic matrix semigroup  $T_{nm}(t)$  with $\sum_n
T_{nm}(t) = 1$, the $H$-theorem holds with the substitutions
$$ 
\phi(\vec{x})\longleftrightarrow p_n \ , \qquad 
\phi_t(\vec{x}) \longleftrightarrow \sum_m T_{nm}(t) p_m
\equiv p_n(t) 
$$
$$
\rho_0({\vec x}) \longleftrightarrow \pi_n = \sum_m
T_{nm}(t) \pi_m  \ , \qquad\  
\int(\;) d\tilde\mu(\vec{x}) \longleftrightarrow \sum_m (\;) \pi_m
$$
{\em The essential condition for the $H$-theorem is
(6.4)}$^{32}$.
To clarify its meaning we go back to diffusion-type Markov semigroups
and notice that if $a_{ij}, \ b_{i} {\in}{\cal D}(Z)$,
use of (5.4), (5.6), (5.7), (6.1) implies
that under the assumptions of theorem 5.1 and for any 
$f\in{\cal D}(Z)$
$$
\ell_\phi(Zf) = \int(Z\phi-H_i\partial_i\phi)f d\tilde{\mu} 
+\int f \phi Z^\dagger \rho_0 d\mu \eqno(6.7)
$$
\n Suppose that $Z^\dagger\rho_0 = 0$. By (6.7) 
$f\to\ell_\phi(Zf)$ is bounded hence it can be uniquely
extended to $C({\Rl}_\infty^n, {\Cx})$ and therefore it belongs to
${\cal D}(Z^\ast)$, and 
$$
(Z^\ast\ell_\phi)(f) = \int Z^\dagger(\rho_0\phi)f d\mu \qquad  
\forall\; f \in {\cal D}(Z) \eqno(6.8)
$$
Since ${\cal D}(Z)$ is dense, $({\tilde Z})^\ast = Z^\ast$, hence
$\ell_\phi\in{\cal D}(({\tilde Z})^\ast)$. Finally  (6.8) can
be applied to $\phi = 1$ and therefore 
$Z^\dagger \rho_0=0$ is equivalent to  $\tilde{\mu}$
being an invariant measure of the Markov process
corresponding to $T_t$. But it
is easily seen that $T_t^\ast \ell_\phi $ is  $s$-continuous, hence 
using theorem 2.1, equation (2.4), we complete the proof of 
\smallskip

\n {\bf Proposition 6.1} If $a_{ij}, b_i$ in (5.3) are in
${\cal D}(Z)$ and $\rho_0$ in (5.4) is annuled by
$$ 
Z^\dagger \equiv \partial_i (a_{ij} \partial_i + (\partial_j a_{ij} - b_i))
\eqno(6.9)
$$
$$ Z^\dagger \rho_0 = 0  \eqno(6.10) $$
then (i) any $\phi\in C^2({\Rl}_{\infty}^n,{\Cx})$ defines an
$\ell_\phi \in{\cal D}({\tilde Z}^\ast)$ via (6.1) and
$$
{d \over dt} T_t^\ast \ell_\phi  =
({\tilde Z})^\ast T_t^\ast \ell_\phi  = Z^\ast T_t^\ast \ell_\phi
\eqno(6.11)
$$
\n (ii) $\tilde{\mu}$ in (5.4$''$) is an invariant measure of
the corresponding Markov process.

\n (iii) If (6.3) holds and $q(t, \vec{x}, {\cdot} )$ is a
$C^2$-function for $t>0$, then (cf.(6.5)) satisfies
$$ 
{\partial\phi_t\over \partial t}=Z^\dagger\phi_t  \eqno(6.12) 
$$
\n {\bf Proof:} The first two assertions have already been
proved. For (6.12), we notice that by (6.2), (6.5),
(6.11),
$$ 
\lim_{t\to t_0}
\int\left({\phi_t(\vec{x})-\phi_{t_0}(\vec{x})\over t-t_0} -
Z^\dagger \phi_{t_0}(\vec{x})\right) A(\vec{x}) d\mu(\vec{x}) = 0 
$$
\n {\em uniformly} in $A\in C_c({\Rl}^n,{\Cx})$. For 
$x_0\in{\Rl}^n$ we take	$f\in C_c({\Rl}^n,{\Cx}) $ nonegative,
 sufficiently differentiable with $f(\vec{x}_0)>0$ and $A_n(\vec{x})$ a
$\delta_{x_0}$-sequence of $C^\infty$-functions. If
$t_n \to t_0$ as $n \to +\infty$, 
applying the above equation to  $A_n(\vec{x}) f(\vec{x})$, written as
$$
\lim_{n \to +\infty} \int f(\vec{x})G_n(\vec{x}) A_m(\vec{x})d\mu(\vec{x}) = 0
$$
we can show from the uniform convergence with respect to $m$
of this double sequence, that $\lim_{n \to +\infty}
G_n(\vec{x}_0) = 0$, i.e. (6.12) holds.\hfill  Q.E.D.
\smallskip

\n {\em Remark:} In the formalism of Refs. 6, 7
$H_i$ in (5.7)  is zero (equation (2.19) of  Ref.7 and the
discussion following it). By (5.7), (6.9),
$(5,4')$ this {\em implies} (6.10). The essential content of
this proposition  is that {\em the existence of an invariant
measure $\tilde{\mu}$, hence of an  $H$-theorem, is 
supplied by the existence of an equilibrium distribution
satisfying (5.4)}. Physically speaking, this may be
interpreted  as the stationarity of a Maxwell-Boltzmann (MB)
distribution for the formal  adjoint equation (6.12), if $H$ in
(5.4$'$) is nonegative.
\smallskip

At this point however it should be noticed that {\em although the existence
of a stationary  solution of (6.12) 
 is not a strong requirement}, on  the contrary, it is a
physically desirable feature of classical kinetic  equations,
{\em  its Lebesgue integrability} (see (5.4)), is a constraint 
{\em violated by quite simple evolution equations and for which a
differential form of  an $H$-theorem is nevertheless easily
proved} (cf.\ (6.14) below and the  discussion following it).
Think for instance of the diffusion equation, even with
variable coefficients, for which $\rho_0(\vec{x})=1$ (see also
appendix 2).
Therefore, one is tempted {\em to relax (5.4) and keep only the
stationarity of $\rho_0$}.  Then, a careful reconsideration of
the proofs of theorem 6.1 and proposition 6.1 shows that {\em minor
modifications} in their hypotheses {\em ensure their validity in
this case as well}. This is the content of
\smallskip

\n {\bf Corollary 6.1 }  With the notation of theorem 6.1 and
proposition 6.1, suppose 
$\rho_0$ is a nonegative, $C^2$-function, bounded at
infinity (or on the boundary of a relatively compact subset of
${\Rl}^n$ - cf. remark to theorem 5.1)
 together with its first derivatives, satisfying
(6.10); $\phi$ in (6.5) has compact support; $h(0)=0$ for $h$ in (6.6).
 
\n Then, (6.12) is valid and (6.4) holds {\em locally}, i.e.
$$
\int q(t,\vec{y},\vec{x}) \, \rho_0(\vec{y}) d\mu (\vec{y}) =
\rho_0 (\vec{x}) \eqno(6.4')
$$
from which the H-theorem follows (cf.(6.6)).
\smallskip

We may use proposition 6.2 to get an instructive, albeit
more special, differential form of the H-theorem : If
$h$ is twice differentiable, then by (6.12), at least formally
$$
{dH \over dt}=\int h'({\tilde \phi}_t)Z^\dagger({\tilde\phi}_t\rho_0)d\mu 
\eqno(6.13)
$$
 where ${\tilde\phi}_t = \phi_t/\rho_0$, $h'(y)=dh/dy$.
Using (5.7), (6.10) directly and in the form 
$\int h({\tilde \phi}_t) Z^\dagger \rho_0 d\mu  = 0 $
we find after a straightforward but lengthy calculation,
$$
{dH \over dt}=-\int \rho_0({\tilde \phi}_t)h''({\tilde \phi}_t)a_{ij}
\partial_i{\tilde \phi}_t \partial_j{\tilde \phi}_t d\mu \eqno(6.14)  
$$
under the assumption
$$
(\rho_0 a_{ij} \partial_jh({\tilde\phi}_t)+
h({\tilde\phi}_t)H_i)|_\infty  = 0 \eqno(6.15) 
$$
(for instance if $h$ is a $C^1$-function with $h(0) = 0$,
e.g. $h(x) = x^2$).

Since $h''\geq 0$, {\em the $H$-theorem holds if and only if
$a_{ij}$ is a nonnegative-definite matrix} (cf. proposition
4.3 of Ref.6; for stochastic matrices see Ref.2 section V.5). 
This shows explicitly the close connection
between {\em conservation of positivity}  (cf. (3.4), (5.3)) and
the {\em existence of $H$-functions}, though theorem 6.1 is more
general. In view of the remark  to proposition 6.1 and apart from
regularity conditions on the coefficients of  the generator
$Z$, equation (5.3), {\em such an operator having a
nonegative-definite second order coefficient}, leads to a
satisfactory kinetic description (c.f. end of section 2) in
the  sense that it {\em defines a Markov semigroup satisfying a
general $H$-theorem,  once the existence of a bounded at
infinity  equilibrium solution (possibly of the MB-type) for
its formal  adjoint has been assured.} As a final
remark, notice that by the above discussion (cf.(6.14)), {\em such a
solution ensures that the Markov semigroup property is 
equivalent to the existence of $H$-functions} (cf.\ in this
respect Refs.33).

As already mentioned, {\em this follows if} $H_i$ = 0,
equation (5.7), which is valid in the formalism
developed in Refs.6, 7 thus proving that the latter
provides a satisfactory kinetic description for open systems
interacting with  large baths at canonical equilibrium.

\vskip 0.5truecm
{\bf Acknowledgement:} This work has been partly supported by the
Euratom Fusion Programm.

 \vfill{\eject}


\centerline{\bf APPENDIX 1}
\bigskip

The characterization of Markov semigroups in terms of Markov
processes is well-known, but proofs are often incomplete or
different because of differences in the definition of a Markov
process. Here we use 
\smallskip

\n {\bf Definition A.1.1} Let $M$ be a compact topological
space, and  $B$ its Borel $\sigma$-algebra. Then a (time
homogeneous) Markov process on $M$ is a function
$p:{\Rl}^+ \times M\times B \tini {\Rl}$ such that 

\n (i) $p(t,x,\cdot)$ is a probability measure

\n (ii) $p(t,\cdot,E)$ is a Borel function

\n (iii) $$ p(t+s,x,E)=\int p(t,x,dy)p(s,y,E) \eqno(A.1.1)$$

\n  For any $f\in C(M,{\Cx})$
 $$T_t f(x)\equiv \int p(t,x,dy)f(y) \eqno(A.1.2)$$
is continuous. The process is called uniformly stochastically
continuous if $^{16}$ 
$$ 
\lim_{p\to 0^+}p(t,x,E)=1 \qquad \mbox{uniformily in $x\in E$} \eqno(A.1.3) 
$$
\smallskip

\n $T_t$ is called a Feller semigroup and
we have the following theorem$^{34, 35}$ used in the proof of theorems 3.1, 4.1
\smallskip

\n {\bf Theorem A.1.1} Uniformly stochastically continuous
Markov processes on $M$ are in 1-1 correspondence with
$s$-continuous Markov semigroups on 
$C(M,{\Cx})$, the latter being equipped with the supremum norm.
\smallskip

\n {\bf Outline of proof:} ($\Rightarrow$) It is easily shown that 
(A.1.2) defines a positivity and normalization preserving  semigroup.
By a standard argument, uniformity of the limit in (A.1.3) implies
s-continuity of $T_t$.

\n ($\Leftarrow$) Let $T_t$ be a Markov semigroup on
$C(M,{\Cx})$. For each $x\in M,\;t\geq 0$
$$
\ell:C(M,{\Cx})\rightarrow {\Cx}:\quad \ell(f)=T_tf(x)
$$
 is a positive (hence bounded) linear functional on
$C(M,{\Cx})$. By (2.6) and theorem 2.1 there exists a unique 
probability measure $p(t,x,\cdot )$ such that  (A.1.2) holds.
(A.1.1) follows from the uniqueness of
$p(t,x,\cdot )$ and the semigroup property of $T_t$. Condition (ii) is proved
by using for every compact or open $E$, a monotonic sequence of
continuous functions $f_n$, converging pointwise to $\chi_E$$^{28}$
to show that  $\lim_{n\to +\infty}T_t f_n(x)=p(t,x,E) $, so that
condition (ii) holds for such $E$. Then taking account of the
regularity of $p$ $^{14}$, this result is extended to all Borel sets.
(A.1.3) is proved by taking for every Borel set 
$E$ and $x\in E$, an $f\in C(M,{\Rl})$ vanishing outside $E$, and
$f(x)>0$, and consider the {\em continuous} function
$$
g(y)=\cases {1, & $f(y)\geq f(x)$ \cr {f(y) \over f(x)}, &
$f(y)<f(x)$ \cr}
$$
for which $ 0<g(y)\leq 1$ for all $y$ such that $f(y)>0$. Then it is
easily shown that $g(x)-T_t g(x)\geq 1-p(t,x,E)$ and (A.1.3) follows
by the s-continuity of $T_t$. \hfill Q.E.D.
\smallskip

\n {\em Remarks:} 
(a) It can be shown$^{34}$ that by the compactness of $M$, pointwise
convergence in (A.1.3) is {\em equivalent} to uniform stochastic
continuity, based on the fact that pointwise convergence of $T_t$ for
$t=0$ is equivalent to w-continuity  for $t=0$.

\n (b) Using (2.3) and the regularity of $p(0,x, \cdot)$, we get
 $p(0,x, E)=\chi_E(x)$ for any Borel set $E$.

\n (c) For every (positive and/or probability) measure $\nu$ defining
 uniquely $\ell \in C^\ast (M,{\Cx})$,  $T_t^\ast\ell$ is uniquely
specified by the (positive and/or probability) measure
$$ 
\nu_t(E)=\int p(t,x,E)d\nu (x) \eqno(A.1.4) 
$$
 
\vskip 0.9truecm
			    
\centerline{\bf APPENDIX 2}
\bigskip

Here we present one-dimensional examples of Markov semigroups, for
which a {\em nonintegrable} equilibrium solution $\rho_0$, satisfying
(6.10) exists, but which nevertheless satisfy the requirements of
corollary 6.1, hence they obey a global H-theorem (for their physical
interepretation see ref.36)
\begin{itemize}

\item[(a)] $ \quad ZA=   (1+x^2) \partial_x^2 A - 
(2\alpha - 1)x \partial_x A  \quad x\in {\Rl}, \ \ \alpha \geq -1/2$

\item[(b)] $ \quad  ZA = x^2 \partial_x^2 A + (1-(2 \alpha - 1)x) \partial_x A
\quad x >0, \ \ \alpha \geq -1/2$

\end{itemize}

\n The corresponding (unnormalized) equilibrium solution $\rho_0$ of
their formal adjoints are respectively 
$$ 
 \rho_0 = (1+x^2)^{-(\alpha + 1/2)}  \ ,  \quad
 \rho_0(x) = x^{-(2 \alpha + 1)} e^{-1/x}  $$
(notice that $H_i$ in (5.7) vanishes identically - cf. remark to
proposition 6.1). It can be shown that for  $-1/2
\leq \alpha \leq 0$, these solutions are not integrable,
although they satisfy the conditions of corollary 6.1, hence the
corresponding semigroups satisfy an H-theorem in the sense of theorem
6.1. Notice that when the phase space is not ${\Rl}_\infty^n$, as in
(b) above, the results of section 6 are still valid provided we
replace any conditions imposed at infinity, with the same conditions
on the boundary of the phase space.



\vspace{1.2truecm}
\centerline{ \sc \bf References}
\footnotesize{
\bigskip 
$^{1}$ H. Risken {\em The Fokker-Planck equation}, Springer, Berlin
(1984).

\smallskip 
$^2$N.G. van Kampen, {\em Stochastic processes in
physics and chemistry}, North Holland, Amsterdam (1992), chs.VIII, IX;

\smallskip 
$^3$R. Balescu, {\em Equilibrium and nonequilibrium
statistical mechanics}, Wiley, New York (1975).

\smallskip
$^4$ The litterature is extensive, see e.g. Ref.3 and
P. R\'{e}sibois and M. de Leener, {\em Classical
kinetic theory of fluids}, Wiley, New York(1977);
R. Kubo, M. Toda and N. Hashitsume, {\em Statistical
Physics}, Vol II, Springer Berlin (1985).

\smallskip 
$^5$C. W. Gardiner {\em Handbook of stochastic methods}, Springer, 
Berlin (1985), ch.3;

\smallskip
$^6$A.P. Grecos and C.Tzanakis, Physica {\bf A151}, 61 (1988).

\smallskip
$^7$C. Tzanakis, Physica {\bf A151}, 90 (1988).

\smallskip
$^8$E.B. Davies, Comm. Math. Phys. {\bf 39,} 91 (1974);
E.B. Davies, Math. Annalen {\bf 219}, 147 (1976);

\smallskip
$^9$ e.g N.G. van Kampen and I. Oppenheim, Physica {\bf A138}, 231 (1986),
section 2 and p.239 eq.(3.5);
V. Rudyak and I. Ershov, Physica {\bf A219}, 351 (1995), eq.(15); 
J. Piasecki and G. Szamel, Physica {\bf A143}, 114 (1987), section 3
eq.(29);
S.A. Aldeman, J. Chem. Phys. {\bf 64}, 124 (1976), eq(3.19);
M. Tokuyama, Physica {\bf A169}, 147 (1990), eq.(4.9a).

\smallskip 
$^{10}$A. Dimakis and C. Tzanakis, J Physics A {\bf 29}, 577 (1996).             

\smallskip
$^{11}$ G. Lindblad, Comm. Math. Phys. {\bf 48}, 119 (1976);
V. Gorini, A. Kossakowski and E.C.G. Sudarshan, J. Math. Phys. {\bf
17}, 821 (1976);
V. Gorini, A. Frigerio, M. Veri, A. Kossakowski and
E.C.G. Sudarshan, Rep. Math. Physics {\bf 13}, 149 (1978).

\smallskip 
$^{12}$There exists an~$A_\infty \in \Cx$, such that for every $\epsilon>0$,
there exists a compact ~$K\subseteq X$ with $|A(x)-A_\infty| <
\epsilon$, $\forall x \in X-K$.
 
\smallskip 
$^{13}$M. Reed and B. Simon,  {\em Methods of modern
mathematical physics I: Functional Analysis},  Academic Press, New
York (1974), chIV.

\smallskip 
$^{14}$W. Rudin,  {\em Real and Complex analysis}, McGraw-Hill,
London (1970), chs. 6, 3, 2, 7.

\smallskip 
$^{15}$K. Yosida, {\em Functional analysis}, Springer, Berlin (1971),
chs.IX, IV, II.

\smallskip 
$^{16}$E.B. Dynkin, {\em Markov Processes}, Vol I, Springer, Berlin
(1965), ch.1.

\smallskip 
$^{17}$E.B. Davies, {\em One-parameter semigroups}, Academic Press, 
London (1981), chs. 1, 2, 7.

\smallskip
$^{18}$R.F. Pawula, Phys. Rev. {\bf 162}, 186 (1967). It contains
Lemma 4.1 specialized to the Kramers-Moyal expansion of the linear
Boltzmann equation.

\smallskip
$^{19}$ J.V. Pul\'{e} and A. Verbeure, J. Math. Phys. {\bf 20}, 2286
(1979).	Lemma 4.1 is shown using a different definition of
dissipativity, {\em a priori} stronger than (3.4).

\smallskip 
$^{20}$E. Nelson, {\em Dynamical theories of Brownian
motion}, Princeton University Press, Princeton (1967); its theorem 5.3
 outlines a proof of theorem 4.1 here.

\smallskip
$^{21}$That this should follow by an
explicit construction of  a counter example (provided by Lemma
4.1) has been pointed to one of us  (C.T.) by E.B. Davies
(private communication).
 
\smallskip
$^{22}$A {\em partial} answer is
provided by the L{\'e}vy-Khinchine formula characterizing Markov semigroups
generated by functions of $\nabla \equiv (\partial_i)$, which
includes (4.2) only when $a_{ij}$, $b_i$ are constants; see e.g.
M. Reed and B. Simon, {\em Methods of modern
mathematical  physics IV:Analysis of operators}, Academic
Press, New York (1978), theorem XIII.53.

\smallskip 
$^{23}$M. Emery, {\em Stochastic calculus on
manifolds}, Springer, Berlin (1989), lemma 6.1.

\smallskip
$^{24}$ $f$ is H\"older continuous with exponent 
$a\in (0,1)$ if $| f(\vec{x})-f(\vec{y})| \leq M{\|
\vec{x}-\vec{y}\|}^a$ for some $M>0$ and $\| \; \|$ is the
Euclidean norm.

\smallskip 
$^{25}$L. Stoica, {\em Local operators and Markov
processes}, Springer Lecture Notes in Mathematics vol. 816, Berlin
(1980).

\smallskip 
$^{26}$R. Courant and D. Hilbert, {\em Methods of mathematical physics}, Vol
II, Wiley, New York (1962), sections IV.2, IV.7;
J.R. Mika and J. Banasiak, {\em Singularly perturbed evolution
equations with applications to kinetic theory}, World Scientific,
Singapore 1995, section 3.7.

\smallskip 
$^{27}$O.A. Oleinik and E.B. Radkevich, {\em Second order
equations with  nonegative characteristic form}, Hogy Nauky,
Moskow (1971), (in Russian), theorem 1.8.2 p.86.

\smallskip 
$^{28}$P.R. Halmos, {\em Measure theory}, Van Nostrand, New York
(1950), sections 55, 36.
 
\smallskip
$^{29}$Sometimes one defines $H(t) = \int h(T_t A)
d\tilde{\mu}$ for the  observables $A$ (see e.g. Ref.15
p.392), though this has no physical interpretation.

\smallskip
$^{30}$If  $(X, \nu)$ is a measure space with
$\nu(X) = N$, and $\phi\in L_\nu^{1}(X)$. Then for any
real-valued convex function $h$ defined on the range of
$\phi$, $ h(\int_X\phi d\nu/N)\leq 
\int_X h\circ\phi d\nu/ N$.
 
\smallskip
$^{31}$Strictly speaking $\int_E (\int
q(t,\vec{y},\vec{x})\rho_0(\vec{y})d\mu(\vec{y})-\rho_0
(\vec{x}))d\mu(\vec{x})=0$ for every Borel set $E$, so
that the integrand is zero $\mu$ - a.e. 
Modifying $q(t,\vec{y},\vec{x})$ on a set of
measure zero we get the result.

\smallskip
$^{32}$By the Banach fixed-point theorem, if (3.11) is relaxed and
 $p(t,\vec{x},{\Rl}^n) < 1$, any invariant measure is unique,  
and if $p(t,\vec{x},{\Rl}^n)\leq a < 1$ its existence is ensured as
well. However (3.11) is essential  in the context of
kinetic theory. Nevertheless, even in this case we do not
know if this invariant measure is  absolutely continuous with
respect to the Lebesgue measure (cf.(5.4$''$)).

\smallskip 
$^{33}$P.M. Alberti and A. Uhlmann, J. Math. Phys.
{\bf 22}, 2345 (1981), theorem 1.2;
P.M. Alberti and A. Uhlmann,  Math. Nachrichten
{\bf 97}, 279 (1980), proposition 5.3.

\smallskip 
$^{34}$J. Lamberti, {\em Stochastic Processes},
Springer, New York (1977), sections 7.6, 7.7.

\smallskip 
$^{35}$J.A. van Casteren, {\em Generators of strongly
continuous semigroups}, Pitnam Publishing Inc., Boston (1985).

\smallskip
$^{36}$ E. Wong, in {\em Proceeding of symposia in applied
Mathematics} vol. XVI, AMS, Providence (1964), p.264.}

\end{document}